# A Kinematic and Kinetic Dataset of Lower Limb Joints During Obstacle Crossing in Healthy Young Adults

Jingwen Huang*, Shucong Yin*, Zhaokai Chen, Hanyang Xu, and Chenglong Fu

Southern University of Science and Technology (SUSTech), Department of Mechanical and Energy Engineering, Shenzhen, China.

**Abstract**

Obstacle crossing is an essential component of human locomotion, particularly for individuals with lower limb amputations who face elevated risks of imbalance and falls. While prior studies have explored this task, they often lack a comprehensive examination of kinematic and kinetic changes throughout the entire gait cycle across varying obstacle heights. This study creates a novel dataset collected from ten healthy adults performing obstacle crossing at four different heights (7.5 cm, 15 cm, 22.5 cm, and 30 cm). Kinematic and kinetic data (angles and torques of hip, knee, and ankle) were recorded and analyzed.

Results indicate that increased obstacle height leads to a longer swing phase and significant increases in both hip and knee joint angles (1.5× and 1.0×, respectively) and torques. In contrast, ankle joint angles and moments exhibited minimal variation across obstacle heights, indicating a relatively consistent movement strategy at the ankle. Furthermore, significant asymmetries were observed between the dominant and non-dominant foot: the dominant foot demonstrated larger hip and knee joint angles and more consistent ankle behavior, reflecting greater coordination. These findings offer valuable biomechanical insights for improving fall prevention strategies and informing the design of assistive devices such as prostheses and exoskeletons.

Corresponding author at: Shenzhen Key Laboratory of Biomimetic Robotics and Intelligent Systems and Guangdong Provincial Key Laboratory of Human Augmentation and Rehabilitation Robotics in Universities, Department of Mechanical and Energy Engineering, Southern University of Science and Technology, Shenzhen, 518055, China.

E-mail address: fucl@sustech.edu.cn (Chenglong Fu)

* contributed equally in this paper (listed in alphabetical order of their family names).

## Introduction

The ability to cross obstacles is essential for human locomotion on complex terrains, including common artificial features such as door thresholds, ground-level water pipes, low fences, and garden edges, as well as natural obstacles like rocks, fallen branches, and uneven debris. Insufficient toe clearance during the swing phase can lead to tripping and serious falls. According to the World Health Organization, about 37.3 million severe fall injuries requiring medical attention occur annually (World Health Organization, 2021), many caused by loss of balance during obstacle negotiation. Such falls result not only in physical harm like fractures but also impose considerable medical and societal burdens (Johnell, 1997; Stevens et al., 2006; Wong et al., 2011). For individuals with lower-limb amputations, the impact is even more critical—especially for those with transfemoral amputations, who have a higher prevalence of falling and fear of falling compared to those with lower-level amputations. A fall may result in physical injury, prosthesis damage, and reduced confidence in walking (Miller et al., 2001). Consequently, developing prosthetic robots capable of autonomously detecting obstacles and adapting their control strategies is vital.

To achieve this, it is common to investigate the relationship between obstacle height and biomechanical parameters using data from healthy individuals. However, no existing dataset fully meets the specific needs of obstacle-crossing analysis. Most current datasets, collected from healthy adults, focus on level walking, treadmill walking, stair ascent/descent, or slope walking (Fukuchi et al., 2017; Hu et al., 2018; Lencioni et al., 2019; McFadyen and Carnahan, 1997; Moore et al., 2015; Moreira et al., 2021; Reznick et al., 2021; Schulte et al., 2023). These studies also examine variables such as walking speed, stair height, and slope angle (Fukuchi et al., 2017; Hu et al., 2018; Lencioni et al., 2019; McFadyen and Carnahan, 1997; Moore et al., 2015; Moreira et al., 2021; Reznick et al., 2021; Schulte et al., 2023). Nevertheless, they do not provide systematic analyses of kinematics and kinetics during obstacle crossing.

Although existing research has provided partial insights into obstacle crossing, these analyses are often incomplete, lacking synchronized joint-level data across the entire gait cycle. Previous work shows that as obstacle height increases, the flexion angles of the hip and knee rise significantly (Chen et al., 2012; McFadyen and Carnahan, 1997), with corresponding increases in joint moments (Austin et al., 1999; Draganich and Kuo, 2004). The trailing limb experiences greater hip and knee moments compared to the leading limb when crossing taller obstacles (Chen and Lu, 2006; Chou and Draganich, 1997). However, these studies often exclude ankle joint analysis and lack complete gait-cycle coverage. Some researchers observed that reduced toe-obstacle clearance increases ankle internal rotation and adduction moments (Chou and Draganich, 1998), while others noted that obstacle width has minimal effect, and height is the dominant factor (Patla and Rietdyk, 1993). Some studies have also noted a relationship between obstacle height and the time required to cross, particularly through increased swing and stance durations (Lim et al., 2023). However, most of these studies focused on individual joints or neglected joint-level analysis altogether, rather than systematically examining the coordinated behavior of all lower limb joints.

In addition to studies focusing on healthy adults, a considerable portion of research has investigated special populations, such as individuals with lower limb amputations and elderly adults. These studies offer valuable insights into compensatory mechanisms and motor adaptation strategies. Amputees often show altered hip and knee kinematics and kinetics to compensate for limb loss (Vrieling et al., 2007), while elderly adults exhibit larger joint angle changes and poorer motor adjustment due to reduced balance and muscle strength (Lu et al., 2006; Park and Lee, 2012). Nevertheless, to support the development of autonomous assistive devices such as prostheses or exoskeletons, biomechanical data from healthy individuals remain essential, as they provide a baseline reference for understanding normal gait.

Although functional asymmetry between the dominant and non-dominant limb is a well-established phenomenon in gait biomechanics, it remains underrepresented in most existing datasets. According to Sadeghi et al. (2000), healthy individuals exhibit functional asymmetry in the lower limbs during gait, with the dominant limb primarily responsible for propulsion, while the non-dominant limb provides stability and control. This asymmetry may extend to obstacle crossing, leading to significant differences between the left and right feet in terms of force patterns, joint range of motion, and spatiotemporal parameters. Such asymmetry should be considered in the design of prosthetics and exoskeletons. However, existing studies lack datasets that comprehensively cover the dominant and non-dominant foot. To address this gap, the present study specifically analyzes bilateral differences to support the development of more adaptive and personalized assistive devices.

To address the limitations above, this study creates a new dataset featuring hip, knee, and ankle joint angles and moments during obstacle crossing in healthy adults. The data focus on sagittal-plane dynamics and include both dominant and non-dominant limbs to capture functional asymmetry. Additionally, the dataset focuses exclusively on the leading limb. This design choice is grounded in the findings of Vrieling et al. (2007), who observed that prosthesis users tend to use the prosthetic limb as the leading leg when a knee joint is present—either in the intact limb or in an active prosthesis. This behavior may stem from the leading limb's ability to adjust in real-time using visual feedback, whereas the trailing limb is partially obstructed from view. Furthermore, when the leading limb is airborne, it is positioned farther from the obstacle due to step length differences, affording more time for foot clearance. By focusing on the leading limb, this study aligns with the practical needs of active prosthetic systems, offering more relevant insights for future assistive device development. The resulting dataset provides a comprehensive view of biomechanical strategies during obstacle negotiation and supports the design of safer, more adaptive wearable robotics.

## Method

### Participants

Ten healthy adults (six males, four females; average age: 20.10 ± 1.14 years, height: 171.69 ± 6.28 cm, weight: 63.85 ± 9.07 kg) participated in the study. Detailed anthropometric data were collected, including the average lower limb length of both legs, measured from the anterior superior iliac spine to the medial malleolus. Dominant foot was determined using the Waterloo Footedness Questionnaire Revised (Yang et al., 2018). Table 1 summarizes participant characteristics.

**Table 1.** Information about ten subjects, including height, weight, age, gender, dominant foot and length of lower limb. *The number within parentheses represents the participants' scores on the WFQ-R questionnaire.

| Number | Age | Gender | Weight (kg) | Height (cm) | Dominant Foot* | Lower Limb Length (cm) |
|---|---|---|---|---|---|---|
| **AB01** | 20 | M | 58 | 170 | Right (6) | 93.8 |
| **AB02** | 20 | M | 80 | 176 | Right (8) | 101.2 |
| **AB03** | 20 | M | 72 | 182 | Right (8) | 102.9 |
| **AB04** | 20 | F | 66 | 163 | Right (8) | 91.4 |
| **AB05** | 18 | F | 60 | 174 | Right (6) | 94.3 |
| **AB06** | 20 | M | 68 | 180 | Right (4) | 99.4 |
| **AB07** | 23 | M | 63 | 173 | Right (8) | 103.2 |
| **AB08** | 20 | F | 56 | 168 | Right (4) | 95.3 |
| **AB09** | 20 | M | 70 | 169 | Right (10) | 93.7 |
| **AB10** | 20 | F | 46 | 162 | Right (8) | 92.1 |

The ratios of obstacle heights (0, 7.5, 15, 22.5, and 30 cm) to limb length were calculated (Table 2). The low standard deviations (<1.5%) across all conditions support the assumption that individual differences in limb length had negligible influence on normalization.

**Table 2.** The ratio of each obstacle height to the participants' lower limb length.

| Obstacle Height | 0 cm | 7.5 cm | 15 cm | 22.5 cm | 30 cm |
|---|---|---|---|---|---|
| **Average** | 0.00% | 8.14 ± 0.35% | 16.27 ± 0.70% | 24.41 ± 1.05% | 32.55 ± 1.40% |

Moreover, according to Draganich and Kuo (2004), walking speed influences the kinematic and kinetic characteristics of the lower limbs. In this study, Participants walked at a self-selected comfortable speed, with the constraint that the walking speed must fall within the range of 0.95–1.10 m/s. The actual walking speed was calculated using the time taken to traverse the 1.75-meter-long force plate. This traversal time—measured using precise gait events from the motion capture system via Cortex software—typically ranged from 1.50 to 1.95 seconds. Trials that fell outside this time range were discarded and repeated to ensure consistency across conditions.

To ensure the accuracy of the data, all participants were provided with comprehensive instructions regarding the experimental procedures and completed the necessary training to familiarize themselves with the obstacle-crossing movements. Before each formal trial condition, participants performed at least five familiarization trials at the corresponding obstacle height. The formal trials only began once the participant demonstrated adequate task familiarity and self-reported confidence in performing the movement. This study received ethical approval from the Ethics Committee of Southern University of Science and Technology **2024PES026**, and all participants provided signed informed consent forms.

**Experimental Setup**

Nineteen reflective markers were placed on the lower limbs and torso following a modified Helen Hayes model (Kerkhoff et al., 2020; Szczerbik and Kalinowska, 2011) (Fig. 1B). During walking, 15 markers remained to enhance comfort and reduce interference, without compromising the accuracy of the experimental results.

A 12-camera motion capture system (Motion Analysis, Santa Rosa, CA, USA) recorded 3D marker trajectories at 120 Hz. Ground reaction forces were captured using two force plates embedded within the deck of a dual-belt instrumented treadmill (Bertec, Columbus, OH, USA). During the experiment, the treadmill belts remained completely stationary and did not move at any time. Participants walked over the treadmill surface as if it were level ground, with the treadmill serving solely as a rigid platform for force measurement. The plates were long enough to cover the entire gait cycle when crossing obstacles, ensuring that complete mechanical data were captured for each step. The force plates captured data at a frequency of 1200 Hz, which was synchronized with the motion capture system to ensure the alignment of temporal and spatial data (Fig. 1A).

**Experimental Process**

The experiment was conducted in a controlled environment to minimize external interference with data collection. Before the formal experiment, participants practiced the obstacle-crossing task, walking along the treadmill and stepping over an obstacle placed mid-path, guided by a marker 25 cm before it. The obstacle had a fixed width of 23 cm (lateral direction) and depth of 15 cm (in the walking direction), with adjustable heights of 7.5 cm, 15 cm, 22.5 cm, and 30 cm. These heights were presented in ascending order to reduce the risk of tripping during early trials. For each height, 20 trials were recorded per leg, totaling 160 trials. Although this fixed order was chosen for safety considerations, we acknowledge that it may introduce order effects such as learning or fatigue. To minimize fatigue effects, participants were provided with adequate rest between conditions. Additionally, five level-walking trials were recorded per participant (Fig. 1A).

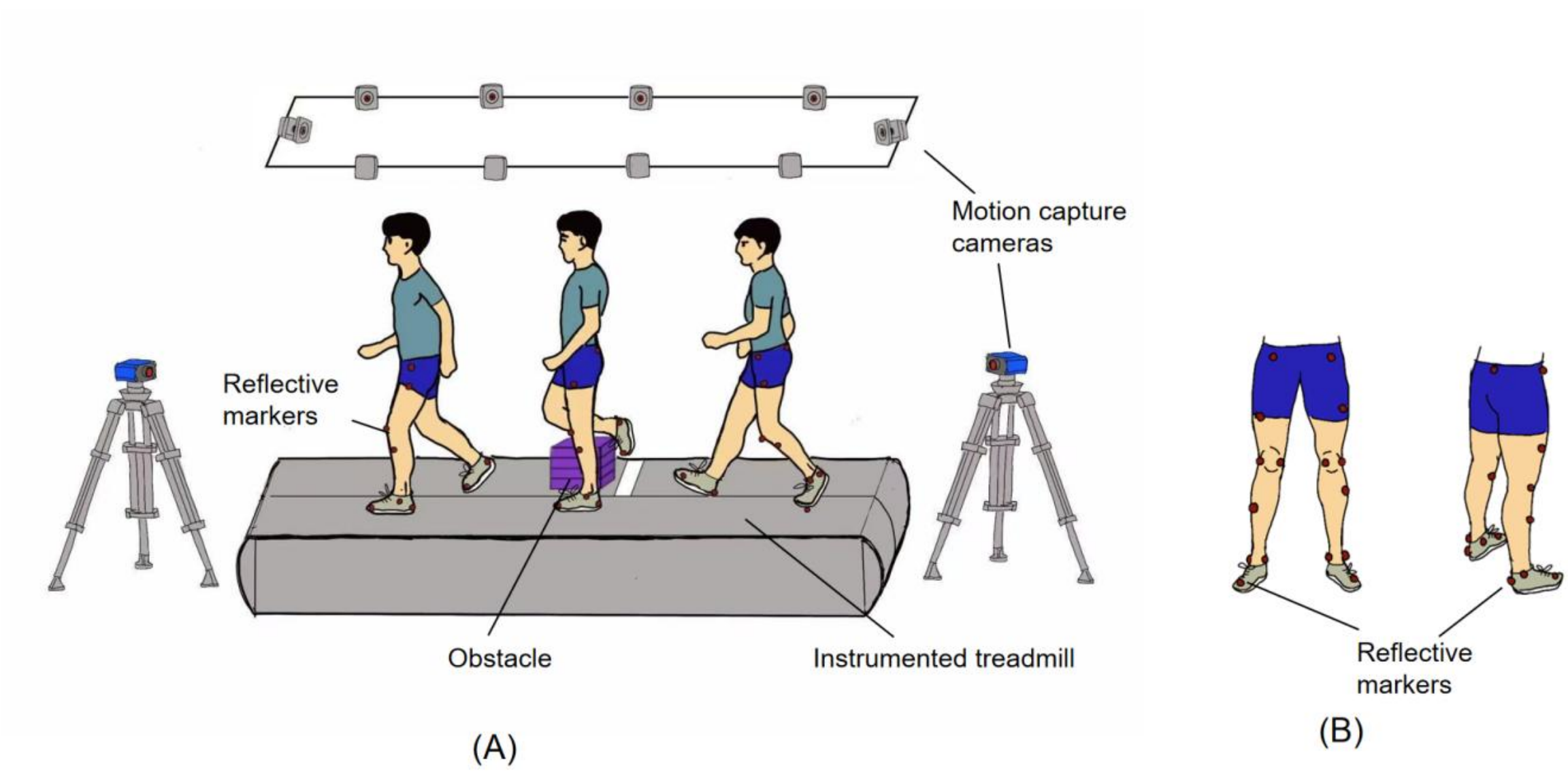

**Fig. 1.** A) The specific obstacle-crossing process of the subject, and the height of the obstacle shown is 30 cm. And the experimental setup: motion capture cameras, an instrumented treadmill, and an obstacle. B) The reflective markers placed on the subject according to the Helen Hayes model.

**Data Processing**

Data processing was conducted using Cortex software (Motion Analysis Corp.). First, raw motion capture data were processed using Cortex's default cubic spline interpolation, which fills gaps up to 10 consecutive frames using natural boundary conditions. A 4th-order low-pass Butterworth filter (cutoff frequency: 6 Hz) was applied to reduce high-frequency noise and to produce smoothed trajectories that closely represent the expected kinematics of human movement during walking. For ground reaction force data, analog force plate signals were low-pass filtered using a 4th-order Butterworth filter with a cutoff frequency of 15 Hz, also based on Cortex's standard configuration. The filtered force data were synchronized with the motion capture data to ensure temporal and spatial alignment. These smoothed data were then imported into Visual 3D (C-Motion), a biomechanics analysis software, for further processing. The motion trajectories of the markers were imported into Visual 3D to calculate kinematic data such as joint angles, angular velocity, velocity, and acceleration of the lower limbs. A skeletal model of the human body was constructed based on the modified Helen Hayes model, which allowed for the calculation of body mass, moment of inertia, and center of mass. Finally, using the 3D ground reaction force data, lower limb joint moments were calculated in Visual 3D through the recursive Newton-Euler algorithm, and joint power was derived from joint moments and joint angular velocities (Ko and Badler, 1996; Koopman et al., 1995).

**Data Availability and Use**

The dataset can be accessed via the website in the Appendix A. The data are stored in Excel format, grouped by participant and leg. Each sheet contains data on the angles, moments, and ground reaction forces of the hip, knee, and ankle joints. The data are organized by obstacle height from low to high for easy analysis using various programming languages, and the data can also be directly opened for inspection. Each data point's gait phase was calculated using linear interpolation between two heel strikes, covering 0% to 100% of the gait cycle. The timing of heel strikes was determined by motion capture data, identified as the moment when the vertical velocity of the heel marker dropped to zero.

Two example scripts are provided to demonstrate how to utilize the dataset. The script Draw_one.py reads the obstacle-crossing data for both legs of a single participant and plots the curves for all the data as well as the average curve. This script can also extract the gait data for a specific participant at a given obstacle height. The second script, Draw_mul.py, can be used to read data for multiple participants and plot the curves for comparison. It is important to note that the joint moments and ground reaction forces in the plotted curves have been normalized by the participant's body mass.

These scripts provide a visualization of the biomechanical changes in the sagittal plane during obstacle crossing. Furthermore, we explored whether the biomechanical data from the sagittal plane could be explained by a linear relationship with obstacle height. Linear regression analysis was performed using custom Python scripts, specifically employing the scipy.stats.linregress() function. At each point in the gait cycle, a linear regression model was fitted to the data to determine the relationship between the biomechanical variables and obstacle height. By calculating the variance explained by the linear model and testing the null hypothesis of zero coefficients ($\alpha = 0.05$), we identified the points in the gait cycle where significant linear relationships between biomechanical variables and obstacle height existed.

**Result**

This section presents the average kinematic and kinetic data for the hip, knee, and ankle joints across the gait cycle during obstacle crossing, normalized by body mass for all ten participants (Fig. 2). Each participant contributed 20 valid trials per leg per obstacle height, and the data were time-normalized to 101 points before averaging. No additional smoothing was applied after averaging. Joint flexion is represented by positive angle values.

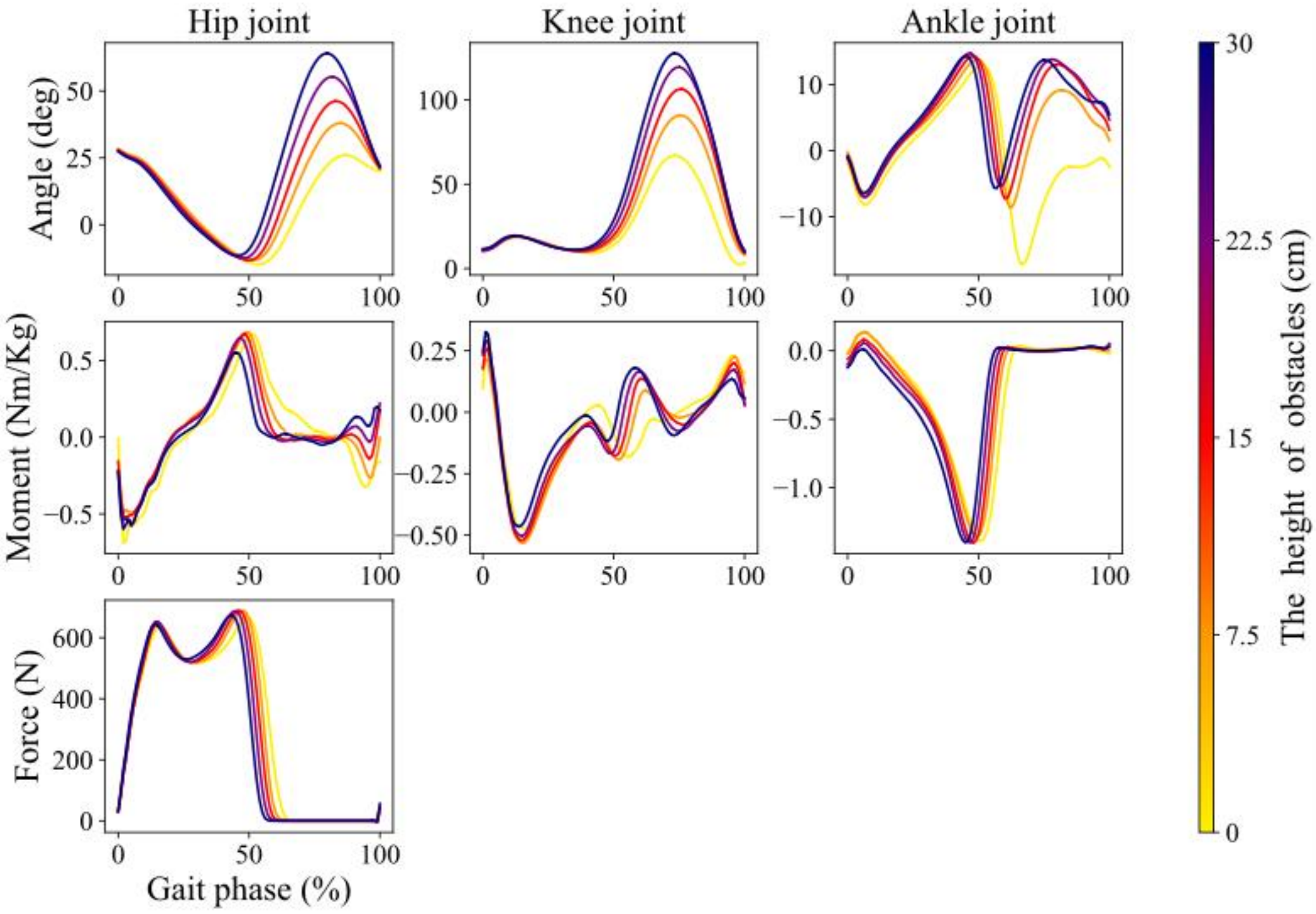

**Fig. 2.** Cross obstacles at different heights. Kinematics (top), moment normalized by subject weight (middle), and ground reaction force (bottom).

As obstacle height increased, the swing phase occupied a larger proportion of the gait cycle. Visualizing the transition points between stance and swing phases across different heights (with shading indicating frequency) revealed that higher obstacles led to earlier swing initiation (Fig. 3), suggesting a potential linear relationship between obstacle height and swing-phase timing. To account for inter-subject variability, data fr om three participants were randomly selected for individual analysis. Transition times varied by approximately 4%, reflecting individual differences. Nevertheless, each partic ipant exhibited a consistent linear trend between obstacle height and swing onse

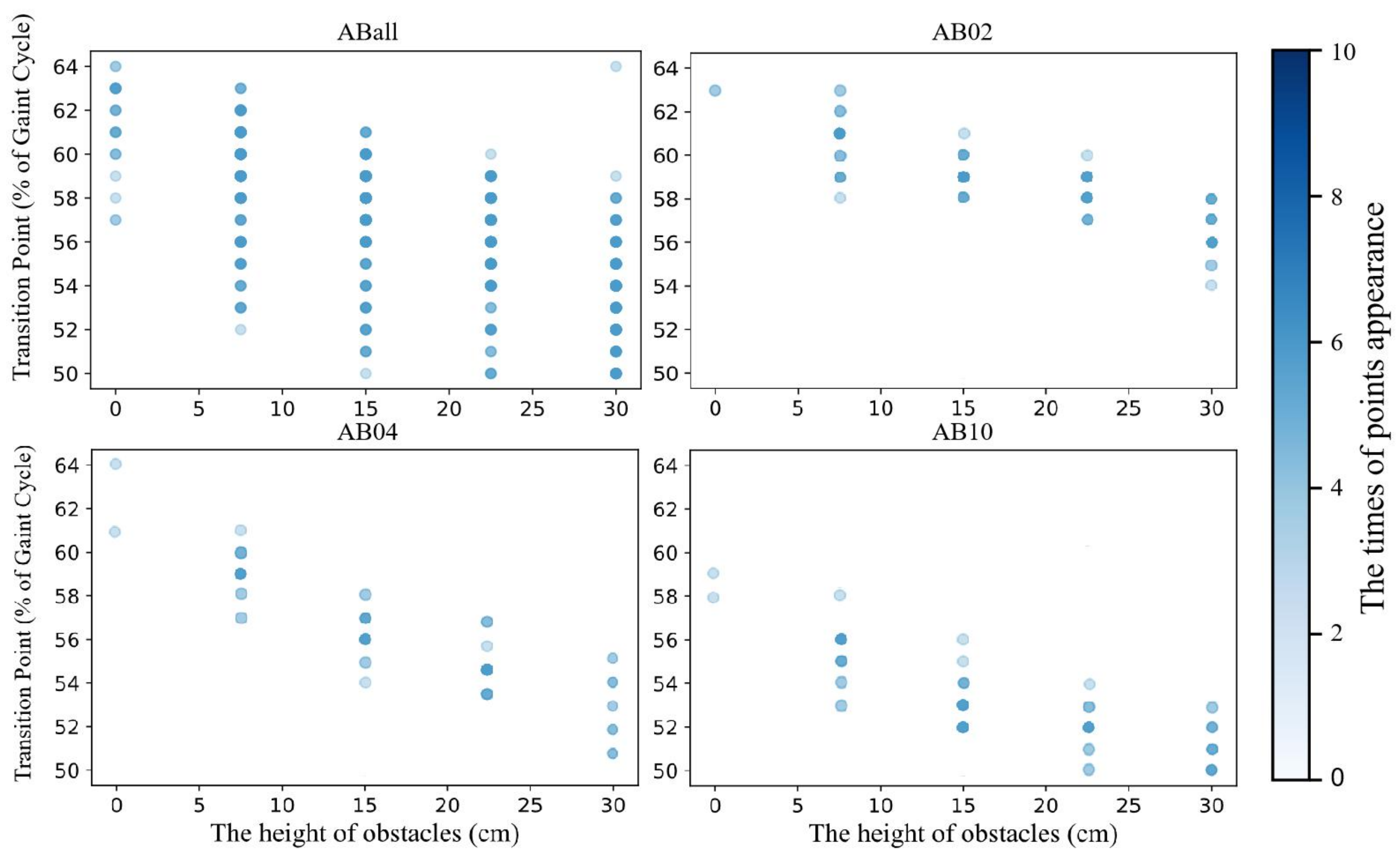

**Fig. 3.** ABall: the data for all subjects; AB02, AB04, and AB10: the data for single s ubjects. The y-axis: the moment when the stance phase is switched to the swing phas e; The x-axis: the change in the height of the obstacle. Color intensity indicates the number of trials in which a given timing occurred, with darker colors representing m ore frequent switching points.

To quantify the relationship between obstacle height and swing-phase timing, a ro bust linear regression model was applied to mitigate the influence of outliers (Evangel opoulou et al., 2016; Huber, 1964; Piradl et al., 2022). The model assigns weights in versely proportional to residuals, ensuring stability in the presence of variability. The f itted equation was $y = -0.2459x + 61.2048$, where x denotes obstacle height and y i ndicates swing-phase transition time. The regression was statistically significant ($p < 0.05$), with tight confidence intervals for the intercept [60.851, 61.558] and slope [–0.26 4, –0.228], indicating high model precision. These results confirm a strong linear relat ionship between obstacle height and the onset of the swing phase, supporting the con clusion that increased obstacle height is associated with earlier leg elevation to facilita te clearance.

Beyond swing-phase timing, we further examined the temporal characteristics of joint parameter peaks. As obstacle height increased, peak timings of joint parameter curves shifted, and is associated with temporal alignment using the ground reaction force (GRF) drop to zero as a marker for swing phase onset. While this adjustment improved alignment for most joints, knee joint angle peaks remained unchanged. Further inspection revealed that the knee angle peaks were already aligned in the raw data, indicating that obstacle height had minimal effect on their timing. As a result, the original knee joint angle curves were retained for analysis (Fig. 4).

To assess the linear relationship between obstacle height and joint-level parameters throughout the gait cycle, a pointwise linear regression was performed at each normalized time point. Given the large number of comparisons, the resulting p-values were corrected using the Benjamini–Hochberg False Discovery Rate (FDR) method ($\alpha = 0.05$) to control for type I errors. Time points exhibiting significant linear trends after FDR correction were visualized as red horizontal bars (Fig. 4). These results revealed that significant linear relationships primarily emerged in the late swing phase, especially in the hip and knee joints. In contrast, the ankle joint exhibited minimal sensitivity to obstacle height, with very few time points reaching statistical significance. The hip and knee moments demonstrated strong linear increases with obstacle height, further reinforcing the conclusion that proximal joints adapt dynamically to support obstacle negotiation, whereas the ankle joint maintains a more stable distal control strategy.

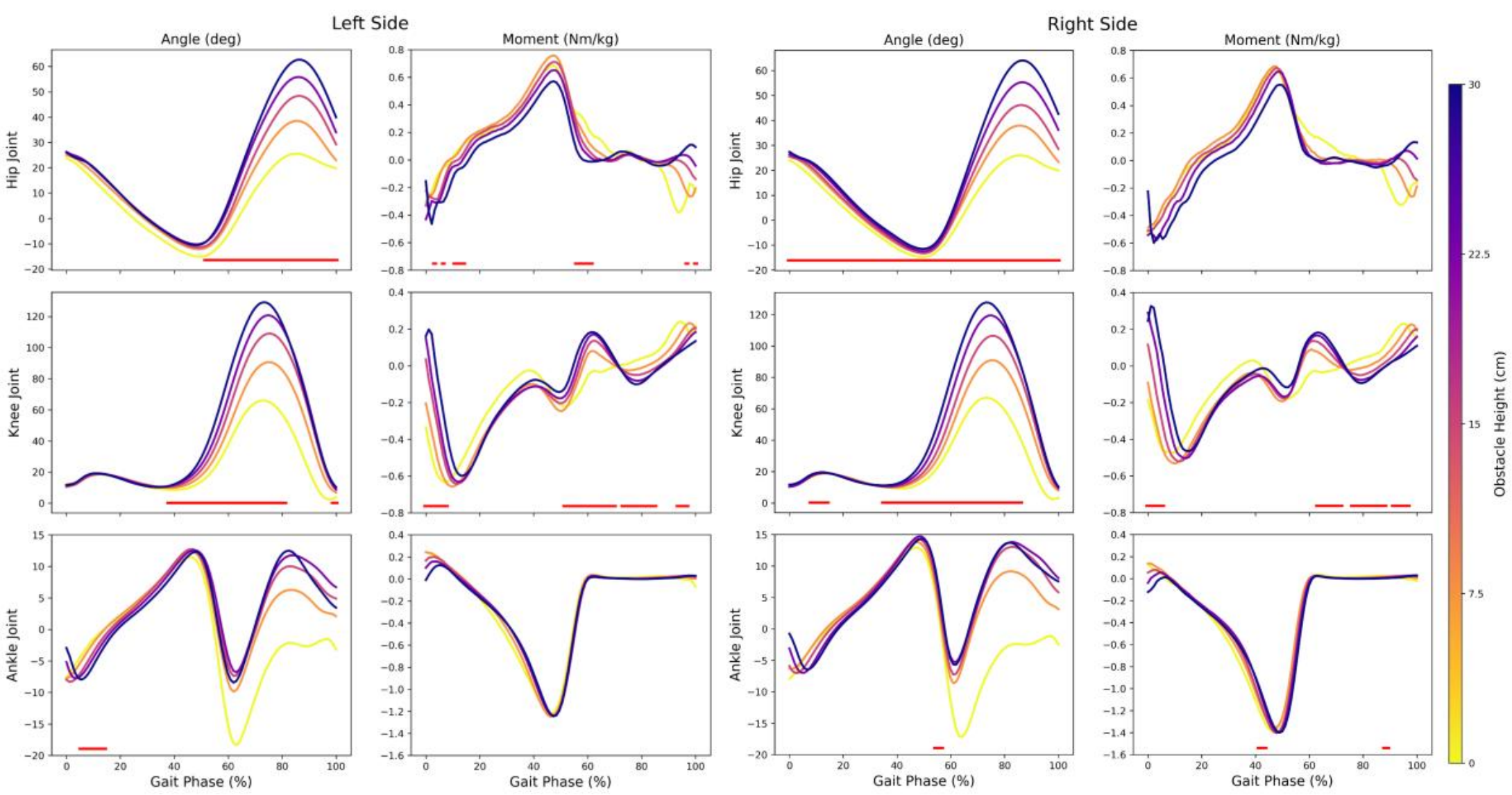

**Fig. 4.** Horizontal red bars denote time points with statistically significant linear relationships to obstacle height after FDR correction ($q < 0.05$).

The maximum dorsiflexion angle during level walking was approximately two to three times greater than that observed during obstacle crossing, as shown in Table 3. At foot contact, the ankle joint was in a plantarflexed position following obstacle crossing, whereas it remained close to neutral during level walking. A linear regression analysis revealed a significant negative correlation between obstacle height and maximum dorsiflexion angle, indicating a progressive reduction in ankle dorsiflexion with increasing obstacle height. However, compared to the substantial increases observed in hip and knee joint angles, the magnitude of change in ankle dorsiflexion was relatively small, suggesting a more constrained role of the ankle in adapting to obstacle height.

**Table 3.** Peak values, linear regression equations for the peaks relative to obstacle height, and their corresponding correlation coefficients.

| **Non-Dominant Joint** | **0 cm** | **7.5 cm** | **15 cm** | **22.5 cm** | **30 cm** | **Equation** | **$R^2$** |
|---|---|---|---|---|---|---|---|
| Hip Joint (Peak) | 25.4 | 38.5 | 48.4 | 55.8 | 62.7 | $y = 1.2253x + 27.7800$ | 0.9903 |
| Knee Joint (Peak) | 66.0 | 90.6 | 109.0 | 120.7 | 129.1 | $y = 2.0840x + 71.8200$ | 0.9782 |
| Ankle Joint(1st peak) | 11.5 | 12.4 | 12.7 | 12.6 | 12.3 | $y = 0.0240x + 11.9400$ | 0.6000 |
| Ankle Joint (2nd peak) | -2.2 | 6.3 | 10.0 | 11.8 | 12.5 | $y = 0.4653x + 0.7000$ | 0.9162 |
| Ankle Joint(valley) | -18.1 | -9.9 | -7.4 | -6.8 | -8.4 | $y = 0.3000x - 14.6200$ | 0.7713 |
| **Dominant Joint** | **0 cm** | **7.5 cm** | **15 cm** | **22.5 cm** | **30 cm** | **Equation** | **$R^2$** |
| Hip Joint (Peak) | 26.0 | 37.9 | 46.2 | 55.3 | 64.0 | $y = 1.2453x + 27.2000$ | 0.9979 |
| Knee Joint (Peak) | 67.0 | 90.9 | 106.4 | 119.4 | 127.6 | $y = 1.9960x + 72.3200$ | 0.9817 |
| Ankle Joint(1st peak) | 12.9 | 13.8 | 14.3 | 14.7 | 14.3 | $y = 0.0493x + 13.2600$ | 0.8444 |
| Ankle Joint (2nd peak) | -2.4 | 9.1 | 13.0 | 13.8 | 13.7 | $y = 0.4920x + 2.0600$ | 0.8463 |
| Ankle Joint(valley) | -17.2 | -8.6 | -7.3 | -5.3 | -5.7 | $y = 0.3507x -14.0800$ | 0.8546 |

Furthermore, there are noticeable differences in the kinematic and kinetic data of the hip, knee, and ankle joints between the left and right sides. Since all participants in this experiment were right-footed, the data from the right leg represent the dominant foot performance, while the data from the left leg represent the non-dominant foot performance.At the ankle joint, the dominant foot exhibited more pronounced plantarflexion, but with significantly less dorsiflexion compared to the non-dominant foot. At the knee joint, the dominant foot demonstrated higher peak knee flexion compared to the non-dominant foot. Similarly, at the hip joint, the dominant foot showed greater hip flexion and higher extension moments during the crossing phase.

**Discussion**

This study examined the effects of obstacle height on the gait cycle and joint parameters. As obstacle height increased, participants initiated the swing phase earlier, and the peak angles of the hip and knee joints increased significantly. In contrast, the ankle joint angles showed only minimal variation across conditions, suggesting that obstacle height primarily affects the proximal joints (hip and knee), prompting earlier swing initiation and increased flexion. Meanwhile, the ankle joint tends to rely on alternative crossing strategies such as external rotation (Chen & Lu, 2006). This shift in control suggests that gait adaptation during obstacle negotiation relies more heavily on proximal mechanisms, with the ankle playing a more supportive and stabilizing role.

The earlier onset of the swing phase with higher obstacles likely provided more time for foot clearance. The increased flexion at the hip and knee supports theories that these joints bear greater mechanical demands during obstacle crossing. These findings highlight the importance of mimicking natural hip and knee behavior in prosthetic and exoskeletal control strategies, supporting more adaptive and biomechanically realistic joint-level actuation.

Additionally, the results reveal significant joint motion differences between the dominant and non-dominant feet, consistent with functional asymmetry in healthy gait. The dominant foot exhibited greater hip and knee flexion, higher extension moments, and more pronounced plantarflexion, but reduced dorsiflexion. These patterns suggest superior neuromuscular control and coordination, enabling smoother motion during obstacle crossing. In contrast, the non-dominant foot showed smaller hip and knee angles but greater ankle dorsiflexion and overall range of motion. Its movement appeared stiffer, likely reflecting reduced motor control confidence. These findings indicate that the dominant foot relies more on proximal adjustments, while the non-dominant foot compensates distally. Incorporating such asymmetry into prosthetic design may enhance biomechanical realism and user performance.

Importantly, unlike prior datasets that focus on isolated gait events or simplified locomotion tasks, our dataset captures the full obstacle-crossing process in the sagittal plane, including continuous kinematic and kinetic variations across five discrete obstacle heights. This comprehensive coverage makes the dataset suitable not only for biomechanical analysis, but also for training and validating prosthetic and exoskeletal control systems. For instance, it may be used to inform rule-based or learning-based control frameworks, support trajectory imitation under variable terrain conditions, or serve as a benchmark for evaluating human-inspired motion planning strategies. Moreover, by quantifying joint-level asymmetries between the dominant and non-dominant limbs, the dataset provides valuable guidance for developing personalized assistive solutions that adapt to individual gait patterns and prosthetic-side constraints. Because the dataset reflects normative crossing strategies in healthy individuals, it can also serve as a training target to guide assistive devices, such as prostheses or exoskeletons, in helping amputees approximate more natural and effective obstacle-crossing movements.

Several limitations of this study should be acknowledged. First, it focused exclusively on obstacle height, without considering other potentially influential factors such as obstacle width, depth, and shape. Second, the analysis was limited to the crossing limb, omitting the contribution of the supporting limb, which plays a critical role in propulsion and balance. Third, obstacle heights were presented in a fixed ascending sequence to reduce the risk of early tripping, which may have introduced order effects associated with motor learning or fatigue. Furthermore, the analysis was confined to sagittal-plane motion. Although this plane captures the primary joint trajectories involved in obstacle clearance and aligns with current prosthetic designs—most of which focus on sagittal-plane actuation—frontal-plane dynamics such as hip abduction/adduction and mediolateral weight shifting are also essential for maintaining dynamic balance during obstacle negotiation. Finally, the sample size was relatively small and limited to healthy young adults, which restricts the generalizability of the findings to broader populations, including older adults or individuals with gait impairments.

Future research should incorporate multi-planar motion capture to enable a more comprehensive analysis of balance control strategies. It should also explore a wider range of obstacle parameters, include bilateral limb analysis, adopt randomized or counterbalanced trial sequences, and recruit larger, more diverse participant cohorts. Such efforts will help deepen our understanding of adaptive gait mechanisms and support the development of more robust and inclusive assistive technologies.

## Conclusion

This study demonstrated that obstacle height significantly influences swing timing and joint parameters, particularly in the hip and knee. It also highlighted functional asymmetries between the dominant and non-dominant limbs. These findings offer theoretical support for improving prosthetic and exoskeleton design through better biomimicry of human gait strategies. Future studies should broaden the analysis by considering additional influencing factors and conducting a more comprehensive examination of both legs. This will further refine gait adjustment theories and enhance their practical applications.

## Appendix A. Supporting information

Supplementary data associated with this article can be found in the online version at https://github.com/Hanyang-Xu/CROSS_OBSTACLE_DATASET.

## Acknowledgements

This work was supported in part by the Shenzhen Science and Technology Program under Grant KCXFZ20230731093401004 and ZDSYS20200811143601004, and in part by the High-Level Special Funds (G03034K003) from the Southern University of Science and Technology, Shenzhen, China. The study was conducted with ethical approval from the Ethics Committee of Southern University of Science and Technology (Approval No. 2024PES026). We would also like to thank all participants for their time and effort during the data collection process.

## CRediT authorship contribution statement

Jingwen Huang: Methodology, Investigation, Data curation, Formal analysis, Visualization, Writing – original draft, Writing – review & editing.

Shucong Yin: Conceptualization, Methodology, Investigation, Data curation, Formal analysis, Visualization, Writing – original draft, Writing – review & editing.

Zhaokai Chen: Methodology, Investigation, Data curation, Validation, Writing – review &

editing.

Hanyang Xu: Investigation, Resources, Writing – review & editing.

Chenglong Fu: Supervision, Project administration, Conceptualization, Funding acquisition, Writing – review & editing.

**Declaration of competing interest**

The authors declare that they have no known competing financial interests or personal relationships that could have appeared to influence the work reported in this paper.